\documentclass[aps,pre,twocolumn]{revtex4}
\usepackage{epsfig}
\usepackage{amsmath}
\usepackage{amsfonts}
\pdfoutput=1
\usepackage{amssymb}
\usepackage{graphicx,natbib}
\newcommand{\be}{\begin{equation}}
\newcommand{\ee}{\end{equation}}
\newtheorem{thee}{theorem}
\newcommand{\bt}{\begin{thee}}
\newcommand{\et}{\end{thee}}
\def\V#1#2{{\mathbf P}_{#1}\{#2\}}
\def\i#1#2{\int\limits_{#1}^{#2}}

\begin{document}
\title{A physical model of nicotinic ACh receptor kinetics}

\author{Ewa Nurowska$^1$\footnote{corresponding author, wrote the
    article,
email: {\tt ewa@fc.units.it}}, Mykola Bratiichuk$^2$\footnote{wrote
Appendix.}, Beata Dworakowska$^1$, Roman J. Nowak$^3$}

\affiliation{$^1$Department of Biophysics, Warsaw University of Life Sciences (SGGW), 02-776 Warsaw, Poland.}
\affiliation{$^2$Institute of Mathematics, Department of Probabilistic and Econometry, The Silesian University of Technology, 44-100 Gliwice, Poland.}
\affiliation{$^3$Institute of Experimental Physics, University of Warsaw, 00-681
  Warsaw, Poland.}

\begin{abstract}
We present a new approach to nicotinic receptor kinetics and a
new model explaining random variabilities in the duration of open
events. The model gives new interpretation on brief and long
receptor openings and predicts (for two identical binding sites) the
presence of three components in the open time distribution: two
brief and a long. We also present the physical model of the receptor
block. This picture naturally and universally explains receptor
desensitization, the phenomenon of central importance in cellular
signaling. The model is based on single-channel experiments
concerning the effects of hydrocortisone (HC) on the kinetics of
control wild-type (WT) and mutated $\alpha$D200Q mouse nicotinic
acetylcholine receptors (nAChRs), expressed in HEK 293 cells. The
appendix contains an original result from probability renewal
theory: a derivation of the probability distribution function for
the duration of a process performed by two independent servers.
\end{abstract}
\date{\today}
\maketitle

\section{INTRODUCTION}
A \emph{generally accepted kinetic model} (GAKM) of ACh receptor
gating assumes that the receptor opens when one or two agonist
molecules get bound to it and shuts before agonist(s) dissociation
(reviewed by Hille (2001)). If only one molecule is bound to the
receptor, it opens for a short time; two bound molecules result in a
long opening of the receptor. It was however noted, that an excess
of brief openings, that appeared in some recordings at high agonist
concentrations, is not consistent with this interpretation
(Colquchoun et al., 1985, Sine et al., 1987 , Hallermann et al.,
2005). Thus, the question of how the
 binding sites contribute to channel gating is still open.

The GAKM assumes that the ACh-AChR binding is formed as a result of
the interaction between the $\pi$-electron systems originating from
the agonist-binding-sites-aromatic-amino-acids, and the cholinum
nitrogen, native to the quaternary ammonium ligands (reviewed by
Arias (2000)).What is usually not discussed is the fact that close
to the ACh binding site there is an electrically noneutral amino
acid, the negatively charged aspartic acid, which may have some
influence on the agonist binding or tracking. To check the role of
this amino acid  for the ACh receptor's kinetics we performed
experiments on the $\alpha$D200Q receptor (Mukhtasimova et al.,
2005) with the negatively charged aspartic acid removed and replaced
by a polar amino acid (glutamine). Previous studies suggested that
the amino acid $\alpha$D200 played a role in the process of receptor
activation, i.e. in the process that starts after a binding of the
agonist (O'Leary et al., 1992, Akk et al., 1996). In our present
studies we go a bit further: we examine the role of this amino acid
in the process of the receptor block.

We studied the receptor's kinetics, in both wild and mutated
versions, to better understand the limits of the validity of the
GAKM discussed above. This model has very severe consequences for
our understanding of the structure of the receptor. Applications of
various blockers, such as the open-channel blockers and the use of
GAKM predicts, for example, the amino acid content of an ion channel
(Leonard et al., 1988). Validity of the models describing how the
receptor is built is thus very strongly dependent on the model of
the receptor's kinetics.

The classical open-channel blockers' theory is based on the
observation that the voltage dependence in the blockers' action
occurs only when the blockers are electrically charged (Neher et
al., 1978, Akk et al., 2003). It turns out, however, that there
exist electrically neutral molecules (such as HC (Bouzat et al.,
1996, Nurowska et al., 2002) and physostigmine (Wachtel, 1993)),
which block the ACh receptor with a very strong voltage dependence.
They prove that the voltage dependency is quite a complex
phenomenon, which is not only a consequence of the molecule charge.
\\Another doubt about the GAKM comes from the fact
that the open-channel blockers action is very often dependent on the
agonist concentration. The occurrence of this dependence is so
common that some authors consider it as the main feature of the
open-channel blockers (Buisson et al., 1998). This is in direct
contradiction with the claim of the open-channel blockers' theory
that the blocker enters into the ionic channel: once the blocker is
in the channel, the agonist concentration should not influence the
block.

In this paper we introduce a model of the receptor blocking which
resolves the problems discussed above. Our model explains how an
electrically neutral blocker (HC) induces a voltage dependent
effect. It also explains the dependence of the blocker action on the
agonist concentration. This model naturally and universally explains receptor desensitization.
Molecular rearrangements causing receptor desensitization were, up to now, poorly understood.
Our model of the receptor block requires a
new kinetic model of the AChR gating. The formulation of this new
kinetic model is also given and its consequences for the kinetic
theory are discussed.

\section{METHODS}

\emph{Cell culture}
 The human kidney cell line HEK 293 was routinely cultured in DMEM,
 supplemented with heat-inactivated foetal calf serum (10\%),
 L-glutamine (4mM), penicillin (100 units$\slash$ml$^{-1}$) and streptomycin
 (100 $\mu$g$\slash$ml$^{-1}$).  Cells were incubated in an atmosphere containing
  5\% CO$_{2}$ at $37\,^{\circ}\mathrm{C}$. One day before transfection, cells were seeded into culture dishes.

\textit{Transfection}. Mouse nAChR subunit ($\alpha$, $\beta$,
$\delta$, $\gamma$) cDNAs in the expression vector pRBG4 (Sine,
1994) and the mito-GFP construct were transiently expressed in HEK
293 cells using LipofectAmine Plus TM Reagent (Life Technologies,
Inc). The WT subunits and the mutated $\alpha$-subunit with the D200
(aspartic acid) to Q (glutamine) substitution were the generous gift
of Prof. Steven Sine (Receptor Biology Lab., Mayo Foundation,
Rochester). Mito-GFP was a kind gift from Prof. Adam Szewczyk
(Nencki Institute, Warsaw). A total of 1  $\mu$g of cDNA per 35 mm
culture dish in the ratio 2:1:1:1:1
($\alpha$:$\beta$:$\delta$:$\gamma$:GFP) was used. The $\alpha$D200Q
contained the mutation on both $\alpha$ subunits.

\textit{Electrophysiology}. Single-channel currents were recorded
from cell-attached patches 1-2 days after transfection (Hamill et
al., 1981). The bath saline (NES) contained (mM): NaCl 140, KCl 2.8,
CaCl2$_2$, MgCl2$_2$, glucose 10, HEPES buffer 10 (pH 7.4). The
pipette contained NES and acetylcholine (ACh, Sigma-Aldrich Co Ltd):
50-500 nM ACh in experiments with the WT receptor and 10-100 $\mu$M
ACh in experiments  with $\alpha$D200Q receptor. HC (Sigma-Aldrich
Co Ltd, 0.02-1 mM) was incorporated in the pipette solution by
dissolving it in NES (Loftsson et al., 2003): the suspension of HC
was heated ($35\,^{\circ}\mathrm{C}$ for about 4 hours) and left
under constant agitation for another 24 h at room temperature.
\\If not otherwise stated, results represent
experiments performed at 60 mV of pipette potential.

\emph{Data Analysis} ACh-induced currents were recorded using an
Axopatch 200B (Axon Instruments, Foster City, CA, USA). The signals
were filtered (2-3 kHz, -3dB) and transferred at 50 kHz, using the
Digidata 1200 interface (Axon Instruments), to a hard disc. Single
currents were analyzed using the pClamp 7 software (Axon
Instruments). Transitions were detected using a threshold-crossing
algorithm, with a threshold for the open and the closed states set
at about 50\% of the mean channel current level. Only recordings
with rarely overlapping openings were analysed and only openings to
the first level were included in the dwell-time distributions. In
case of stable patches, after performing recordings at different
membrane potentials, we performed the second registration at +60 mV
of pipette potential starting about 15-20 min after the patch
formation. The receptor kinetics recorded at +60 mV immediately
after the patch formation and after 15 min were compared. We did not
observe any significant changes neither in time constants nor in
current amplitude. In most of experiments the dwell-time
distributions were constructed from more than 1000 events, however,
in few cases from more than 300 events. Open and blocked time
constants  $\tau_{open}$  and  $\tau_{blocked}$ of the open and
closed time distributions were fitted by the method of maximum
likelihood using the PSTAT program (pClamp7) with a probability
density function being the sum of \textit{n} exponential terms, i.e.
$$ f(t)=\sum\limits_{i=1}^n a_{i}\frac{1}{\tau_{i}}\exp (-\frac{t}{\tau_{i}}) $$
where $a_{i}$ represents the area of the i-th component. The open
time distribution of the WT receptor was fitted with 2 components
(\textit{n} = 2); $\tau_{open}$ corresponds to the long component of
the open time distribution. The open time distribution of the mutant
$\alpha$D200Q receptor was fitted with 1 component only (\textit{n}
= 1). $\tau_{blocked}$ corresponds to the component that is present
in the closed time distribution of the blocked receptor only, and
such, its area increases with HC concentration. It should be also
mentioned that due to the resolution of our single channel recording
(0.110 -0.180 ms), very brief events were probably undetected. This,
however, should not affect the estimation of the duration of the
blocked events ($\tau_{blocked}$). They appear only within the
bursts having at least two openings. It was shown previously
(Nurowska et al., 2002) that in such bursts brief openings are
absent. Since brief openings exist only as isolated openings, they
can not affect the closed periods within a burst.

\textit{Statistical analysis}. Data are given as mean $\pm$ SEM. The
unpaired Student's \textit{t}-test was used to determine the
statistical significance. If not otherwise stated, results are
pooled for different ACh concentrations and number $\it{n}$ of samples is given in parentheses.
\section{RESULTS}
The inclusion of ACh, in the patch pipette, induced single-channel
currents in both types of cells expressing WT receptors or mutated
$\alpha$D200Q nACh receptors (Fig. 1). However, the activity of the
$\alpha$D200Q receptor was visible only when the ACh concentration
was approaching micromoles. Below this concentration of ACh the
activity of the $\alpha$D200Q receptor was negligible. For this
reason, in experiments with the $\alpha$D200Q receptor, we increased
the agonist concentration up to 10-100 $\mu$M. This activated the
receptor, but contrary to the standard behaviour of the receptor in
high concentration of ACh, did not cause receptor desensitization
(Fig. 1A). Actually only one component was present in the closed time
distributions which time constant decreased with the agonist concentration increase.
A second component in the closed time distribution appeared only after adding the hormone.

The single-channel \textit{I-V} relationship gave a slope
conductance of 33.4 $\pm$ 0.8 pS (\textit{n} = 9) for the WT
receptor, and 32.0 $\pm$ 0.8 pS (\textit{n} = 17) for the
$\alpha$D200Q receptor, suggesting similar pore conductance
properties. The mean open-channel lifetime (i.e. the mean time
constant of the main component, $\tau_{open}$) of the WT receptor
was 6.90 $\pm$ 0.78 ms (50-500 nM ACh) and did not depend on agonist
concentration. The $\tau_{open}$ of the $\alpha$D200Q receptor
increased with the agonist concentration increase, being 0.81 $\pm$
0.22 ms for 10 $\mu$M ACh and 1.57 $\pm$ 0.46 ms for 100 $\mu$M ACh
(\textit{P}$ < $ 0.0001, two-tailed \textit{t}-test). Even so,
$\tau_{open}$ of the mutated receptor was significantly smaller than
$\tau_{open}$ of the WT receptor, in the presence of 50-500 nM ACh
(\textit{P}$ < $ 0.001, two-tailed \textit{t}-test; Fig. 1B). The
voltage-dependency of $\tau_{open}$ for the mutated receptor was,
nevertheless, maintained (Fig.1B).

In the presence of 1 mM HC in the patch pipette, there was no
significant effect on the single-channel \textit{I-V} slope
conductance, which was 32.0 $\pm$ 0.5 pS (\textit{n} = 10) for the
WT receptor and 31.6 $\pm$ 0.5 pS (\textit{n} = 12) for the
$\alpha$D200Q receptor. In both types of cells, HC decreased
$\tau_{open}$ in a concentration-dependent manner (Fig. 2), and
produced a characteristic burst-opening effect. When HC blocked the
WT receptor, $\tau_{open}$ did not change with change of the patch
membrane potential, or changed very weakly (Fig. 3). However, when
HC blocked the mutated receptor, $\tau_{open}$ increased with
increasing membrane hyperpolarization (Fig. 3). Interestingly, the
blocked events induced by HC were voltage-dependent in the WT
receptor, but we did not notice any obvious voltage-dependency of
the mean time constants of the blocked events ($\tau_{blocked}$) in
the mutated receptor (Fig. 4). For all HC concentrations
$\tau_{blocked}$ of the $\alpha$D200Q receptor were significantly
shorter than $\tau_{blocked}$ of the WT receptor (Fig. 5A;
\textit{P}$ < $ 0.001). This decrease in the $\tau_{blocked}$ for
the mutated receptor was not induced by the high agonist
concentration, since our experiments with different agonist
concentrations and the $\alpha$D200Q receptor excluded this
possibility (Fig. 5B). In both types of receptors,
$\tau_{blocked}$ depended on the HC concentration (Fig. 5A).
\section{DISCUSSION}

The main conclusion of our experiments is that $1/\tau_{block}$
decreases linearly with the HC concentration growth (Fig. 5A). This
means that the kinetic constant describing
  the passage from a \textit{blocked} to an \textit{open} state is given by
 $a-k [HC]$ where $a$ and $k$ are $[HC]$-independent
  constants. We thus have
$$open\stackrel{a-k[HC]}{\longleftarrow}{block}.$$ This
suggests that the state \textit{blocked} is a complex state
 consisting of two substates $B_1$ and $B_2$,
  between which there exists a passage $B_1\to B_2$ in the direction
  opposite to the direction
$blocked\to open$ with a kinetic constant equal to $k$[HC]:
\be \label{sch} \ee
\begin{center}
\includegraphics[width=2.5cm]{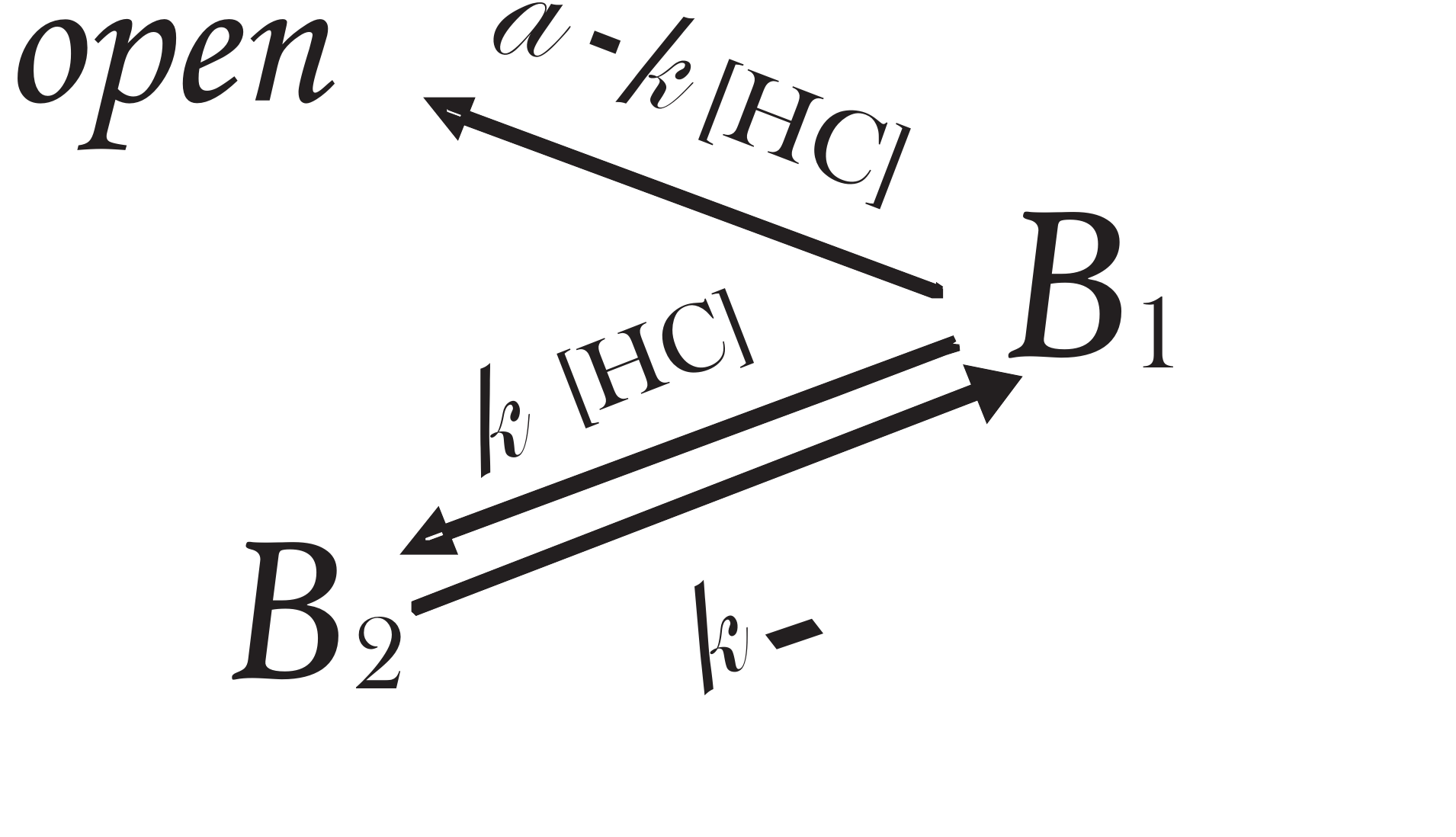}
\end{center}

Here $k_-$ denotes the dissociation constant of the blocker. It
follows that the passage $B_1\to open$ presented by this scheme is
not connected with the dissociation of the blocker; the blocker
dissociates when passing from $B_2$ to $B_1$, thus the state $B_1$
is a state without a bound blocker.

The complex scheme (\ref{sch}) has
  a kinetic constant corresponding to the passage $blocked\to
  open$ given by
\be
k_{open}=\frac{a-k[HC]}{1+\frac{k[HC]}{k_-}}. \label{gw}
\ee
Note that formula (\ref{gw}) becomes $a-k [HC]$ in the limit
  $k_->>k[HC]$. This limit corresponds to the situation when the state
  $B_2$ is very much shorter in the dwell-time than the state $B_1$.

A consequence of the postulated kinetic model described by scheme
(\ref{sch}) is that state $B_1$ is very particular: the receptor
leaves state $B_1$ with the same frequency $a$ independently of the
presence of the blocker. Only after leaving this state the receptor
`decides' to pass to precisely one of the two possible states
\textit{open} and $B_2$. Thus, the two states \textit{open} and
$B_2$, which can not be occupied simultaneously, originate from the
same event: the abandonment of state $B_1$ by the receptor. This
means that the rough scheme (\ref{sch}) describing the kinetics of
the receptor is more precisely represented by: \be \label{sxh} \ee
\begin{center}
\includegraphics[width=2.5cm]{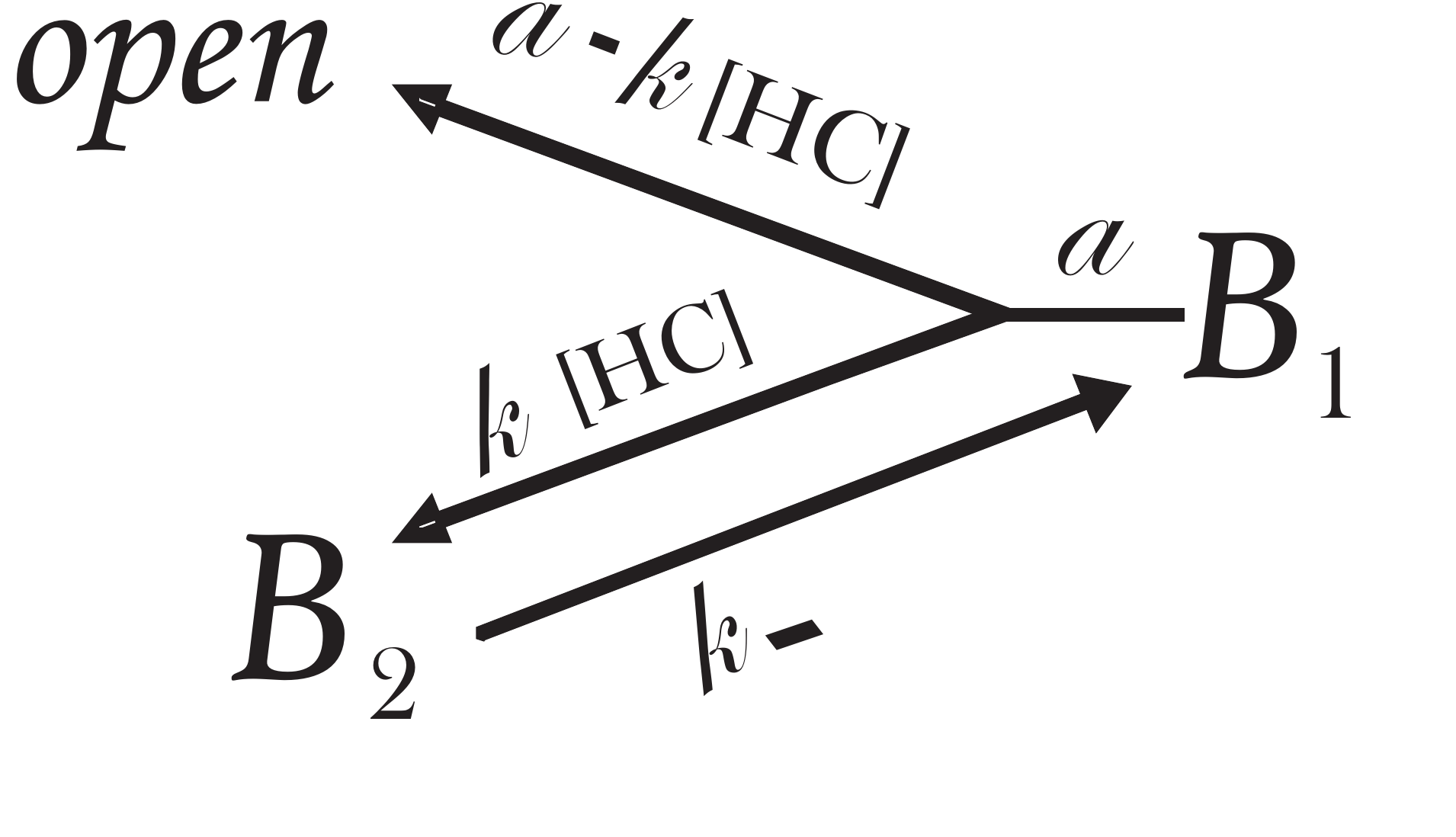}
\end{center}

We further postulate that the state $B_1$ is unliganded. This in
particular means  that the opening of the receptor: $B_1\to open$ is
caused by a binding of the agonist to the binding (activating) site.
It also means that, in an open receptor, the agonist is not
permanently attached to the binding site. It switches between the
state when it is bound, the states when it is off the binding site,
again the state in which it is bound, and so on. The existence of
two binding sites prevents the receptor from being shut: in moments
when the agonist leaves one binding site the receptor will still be
in an open state, provided that another agonist molecule is bound in
the second binding site. The same is true when the second agonist
molecule leaves the binding site: as only when the first agonist
molecule is bound, will the receptor be opened. We assume that after
leaving the binding site the agonist eventually comes back, due to
the attractive electric force associated with the negative charge of
the close-to-the-binding-site-amino acid  D200. This assumption was
confirmed by our experiments: we performed a series of experiments
with a mutated $\alpha$D200Q receptor in which the amino acid  D200
was replaced by a polar amino acid Q (with much smaller negative
charge). As we expected, in such a mutated receptor the openings
were much shorter than the openings of the wild type receptor (Fig. 1B).

These considerations enable us to formulate the following
`oscillating' model of the receptor opening:
\begin{itemize}
\item
The receptor has two sites, say $A$ and $B$, at which the agonist
(also called the activator) can be bound. It opens immediately after
one molecule of the activator is bound to $A$ or $B$, and stays open
as long as \emph{at least one} of the binding sites $A$ or $B$ is
occupied. The activators dissociate from $A$ and $B$ after some
time, $t_A$ and $t_B$, respectively. These times have the same
exponential probability distribution with parameter ${\lambda > 0}$.
\item Let us assume that the receptor opens because of an agonist
binding at $A$. Then the agonist stays
  bound to the receptor at $A$ for time $t_A$. We also assume
  that the second binding site binds the second agonist's molecule
  after some time (which depends on agonist concentration),
  say $\eta$, measured from the moment of the
  opening of the receptor. $\eta$ is a random variable which, if it is
  greater than $t_A$, the opening time of the receptor is simply $t_A$.
\item If $\eta \leq t_A$ the receptor is still opened, but now with \emph{both} $A$ and $B$
occupied. In such situation each of the two agonists may dissociate
independently after the respective times $t_A$ and $t_B$. Each of
them returns to its binding site after time $t_x$, provided that the
other agonist is still bound. The time $t_x$ is a random variable
having the exponential distribution with the parameter ${\mu > 0}$, which is the
same for both agonist molecules.
\item The situation in which the molecules from $A$ and $B$ leave
and return from and to their binding sites repeats, until the moment
in which both binding sites are empty. In such situation the
receptor shuts down.
\end{itemize}

The above assumptions about the receptor's kinetics suffice to
derive an explicit formula for the probability $\V{}{opentime>t}$
of the receptor's open time. This original
derivation is given in the Appendix.

Assuming that the shift time $\eta$ is a random variable with
probability distribution $F(x)$ not depending on $t_A$, $t_B$ and
$\mu$, and denoting  $\tau$=opentime, we obtain the formula for $\V{}{\tau>t}$
 given in the Theorem 1 in the Appendix (expression
\ref{pp}).

We emphasize that the \emph{analytic} expression (\ref{pp}) for the
open time distribution is in total agreement with Monte Carlo
simulations of the open time, assuming that the receptor obeys the
proposed model and has the shift time distribution $F(x)=1-{\rm
e}^{-\nu x}$ (see Fig.6).

The formula (\ref{pp}) is a superposition of three exponential
functions. The first one, which has the time constant -$1/s_1$, we attribute
to the distribution of long openings. The brief openings consist
of two closely spaced components with the corresponding time
   constants equal to: -$1/s_2$ and  $1/(\nu+\lambda)$. Note, that
   neither
    of them has a time constant equal  to the mean time a single agonist
    molecule stays in the activating site, contrary to
what is often assumed for brief openings.
    Note also that three components in the open time distribution
   were already observed in high resolution
  (Hallermann et al., 2005, Parzefall et al., 1998) and classical (Hatton et al., 2003) single channel recordings.
   The formula (\ref{pp}) fitted by the minimum $\chi^2$ method
  was in perfect agreement  with the open time distributions
measured for WT and  $\alpha$D200Q receptors in all 38 recordings.
  Examples of the comparison of the theoretical distribution
(\ref{pp}) with experimentally obtained distributions of the open
events for both receptors are shown on Fig.6. The mean values of the
fitting parameters $\lambda$ and $\mu$ did not depend on agonist
concentration and were: $\lambda$ = 1.03$\pm$0.19 ms$^{-1}$,  $\mu$
= 19.0$\pm$4.6 ms$^{-1}$ for the WT receptor ($\it{n}$ = 17) and
$\lambda$ = 1.35$\pm$0.11 ms$^{-1}$,  $\mu$ = 3.30$\pm$0.76
ms$^{-1}$  for the $\alpha$D200Q receptor ($\it{n}$ = 21). Similar
values of the $\lambda$ parameter suggest that the times the agonist
stays in the activating site ($t_A$ and $t_B$) are not affected by
the presence of the amino acid D. However, since the $\mu$ parameter
obtained for the WT receptor differs significantly ($\it{P}<
$0.0001) from the same parameter obtained for the mutated receptor,
we suggest that when closed to the activating site, the agonist
stays in the electric field of the amino acid D. This refers to the
open receptor. Mean parameters $\nu$ obtained for both receptors do
not differ significantly (for the WT receptor $\nu$ = 3.64$\pm$1.20
ms$^{-1}$ and for the $\alpha$D200Q receptor $\nu$ = 1.64$\pm$0.47
ms$^{-1}$), despite the different agonist concentration used in the
experiments. Thus, $\nu$ parameter depends not only on the agonist
concentration, but also on the presence of the amino acid D.

We now pass to the discussion of the receptors's blocking mechanism.

We assume that the unblocked receptor acts according to the kinetic
model presented above. This model admits a shortening of the opening
time in the situation when one of the binding sites is unoccupied.
In such situation the returning-to-its-binding-site-agonist may
encounter a molecule that blocks its entrance to the site (without
really binding to this site), resulting in turn in the closure of
the receptor. We postulate that the molecule responsible for such a
block is HC.

To present the detailed blocking mechanism we first explain what
happens in the blocked receptor with the agonist molecules which
activated it. The observed existence of bursts suggests that these
molecules do not return to the extracellular solution. Indeed, if
the receptor in the $B_1$ state was activated by binding the agonist
molecules from the extracellular solution, then $\tau_{blocked}$
would \emph{decrease} with the agonist concentration. Since this is
not observed (Fig. 5B), we conclude that the agonist needed for the
transition $B_1\to open$ comes from the region which is not
accessible to other molecules of the activator. Since it is widely
accepted (Miyazawa et al., 1999) that the receptor's activating
sites are located in cavities accessed through narrow tunnels, we
conclude that even after the dissociation the activating molecules
remain in their tunnel-cavities. Since these cavities are of the
correct size and shape to accommodate one ACh molecule (Miyazawa et
al., 1999) no other agonist molecule could bind to its binding site
until the ACh molecule leaves the tunnel-cavity. Note that the
agonist staying in the cavity-tunnel experiences the electrostatic
repulsion from other agonist molecules trying to enter the tunnel.
It forces the agonist to stay at the bottom of the tunnel. If the
agonist binding site is not located there, then $\tau_{blocked}$
should \emph{increase} with the ACh concentration. Such a behavior
of $\tau_{blocked}$ was observed in our experiments with receptor
$\alpha$D200Q ($P < $0.001, Fig. 5B). We thus postulate that the
agonist binding site is situated in another place than the end of
the tunnel (Fig. 7).

The main difference between state $B_1$ and the closed state is that
in state $B_1$, both tunnels associated with the activating sites
are occupied by the corresponding agonist molecules, which at every
moment may be bound to the activating sites (recall that state $B_1$
is a state without a blocker). Surprisingly state $B_1$ (without a
bound blocker) lasts longer than the blocked state $B_2$. This is
again caused by the repulsive action of the outside agonists
molecules which are close to the entrance to the tunnel. They
prevent the agonist in the tunnel from reaching the binding site.
Since these molecules are not present at the entrance to the tunnel
all of the time, there are moments when the entrance to the tunnel
is free. This occurs with frequency $a$. At such moments the agonist
within the tunnel can approach the entrance and bind to the site,
opening the receptor again. Moments when the entrance to the tunnel
is not occupied by the outside agonist molecules can be also used by
the blocker to enter the tunnel. The blocker, as electrically
neutral, does not feel the repulsive potential of the agonist within
the tunnel. Entering the tunnel, it can freely penetrate it up to
the agonist binding site. Once there, it prevents the inside agonist
from binding. Note that the presence of the blocker does not
decrease the duration of the $B_1$ state: the blocker has a chance
to enter the tunnel only when an agonist may return to the binding
site, thus only when the receptor leaves state $B_1$. This agrees
with scheme (\ref{sxh}).

Since external molecules repulse the inside tunnel-cavity agonist
from the activating site, it is clear that $\tau_{block}$ is longer
in a receptor in which the entrance to the tunnel is more frequently
occupied by the agonist. The measured $\tau_{blocked}$ (Fig. 5A)
indicates that such situation occurs in the WT receptor (despite of
the lower concentration of the agonist!), which thus must have
agonist's molecules at the entrance to the tunnel more often than
the $\alpha$D200Q receptor. We believe that the reason for this
difference is the negative charge of the amino acid $D200$, which
most probably attracts positively charged molecules of the activator
towards the entrance of the tunnel in the WT receptor. This also
explains why the probability of the opening of the WT receptor is
higher than the probability of the opening of the $\alpha$D200Q
receptor.

This justifies our next postulate, that the location of amino acid
$\alpha 200$ is different in an open and a closed receptor: if the
receptor is open, the location of amino acid $\alpha 200$ is close
to the activating site \emph{inside} the cavity-tunnel; if the
receptor is closed, $\alpha 200$ is located \emph{outside} the
cavity-tunnel, but still close to its entrance.

Summing up: it is the HC molecule that initiates the blockage of the
receptor. At the initial stage of the block ACh molecules do not act
as blockers, since the open receptor does not attract them from the
extracellular region toward the tunnel-cavity where the block
occurs. Once the receptor is in the blocked state the situation
changes: the closed receptor exposes the negative charge to the
outside of the tunnel, starting the process of attracting the
extracellular molecules towards the tunnel-cavity. Thus, in the
process of blocking, both ACh and HC participate. On the one hand
the ACh molecule, located at the entrance to the tunnel-cavity,
repulses the agonist trapped in the tunnel from the activating site,
on the other hand the HC molecule, when the tunnel entrance is
empty, enters into it, preventing other molecules from reaching the
activating site.
If the location of the amino acid $D$ was changing with the membrane
potential, then the action of HC would be unaffected (HC is an
uncharged molecule). But it would alter the action of ACh. We
postulate that the amino acid $D$ location \emph{is} voltage
dependent. This implies the voltage dependency of $\tau_{block}$ in
the WT receptor. In $\alpha$D200Q the situation is different. The
frequency of the ACh approaches to the tunnel cavity mainly depends
on the ACh concentration, and not on the attraction of the amino
acid $\alpha$200, which in the case of the amino acid Q is weak.
Thus in the mutated receptor $\tau_{blocked}$ has no significant
voltage dependence (Fig. 4).

Finally we emphasize, that although the state $B_1$ is very
peculiar, we believe that it occurs very commonly. For this state to
exist the blocker is not needed. It is a closed state, which differs
from the usual closed state by the presence of unbound agonist
molecules in the tunnel-cavities in both subunits. If such a state
randomly emerges (e.g. in high concentration of the agonist) then,
if the agonist concentration remains high, exiting from this state
would be very difficult: the highly concentrated activator molecules
will be present at the entrance to the tunnel-cavity repulsing the
agonist from the binding site. We associate our state $B_1$ with a
particular closed state of the receptor which is reported to be
observed only in high agonist concentration and to be characterized
by a very long duration. Thus, state $B_1$ corresponds to the
desensitization of the receptor. That in the process of this
desensitization amino acid $D200$ plays a role is confirmed by our
experiments, in which such desensitization was not present in the
receptor $\alpha$D200Q, even if we went to agonist's concentration
as high as 0.1 mM.


Experiments with other blockers and $\alpha$D200 mutants are needed
to confirm the role of $\alpha$D200 amino acid in mechanisms of the
receptor block and desensitization. A presence of charged amino
acids close to the agonist binding pockets may be in this family of
receptors (shared by the nicotinic, the GABA$_A$ and the 5-HT$_3$
receptors) the common feature (Schreiter et al., 2003, Hartvig et
al., 2000), making the proposed model of the receptor
desensitization universal for other ligand-gated ion channels.

\section{Appendix}
Let us consider a  system formed by two identical servers (say A and
B ) working independently and let $\xi_1,\xi_2$ stand for the time
of their trouble-free operation. It is supposed that
$\V{}{\xi_i>t}=e^{-\lambda t}, \ \lambda>0,\ t \ge 0, \ i=1,2.$ At
the initial time $t=0$ both servers are in working condition and,
say, server A starts to work. The second server starts to work with
some delay  $\zeta\ge 0 $ having a distribution $\V{}{\zeta<t}=
F(t), \ t\ge 0.$ In the case of a breakdown of any server its
repairing starts immediately and it lasts  $\eta_1, \eta_2$
time-unites respectively. It is supposed that
$\V{}{\eta_i>t}=e^{-\mu t},\ \mu>0 ,\ t \ge 0 i=1,2.$ After the
repair is finished the server starts to work again. The above
mentioned random variables are all supposed to be independent.

The breakdown of the whole system occurs if both servers are not in
working conditions  (i.e., they are being repaired).

Let  $\tau$ denote the time at which the system breaks down for
the first time. We are interested in finding $ \V{}{\tau>t}.$ This is
given in the following theorem.

{\bf Theorem 1}\\
\begin{align} \label{5} \V{}{\tau>t } =
\i{0}{t}e^{-\lambda u}\Psi(t-u)dF(u) +(1-F(t))e^{-\lambda t},
\end{align}
where
\begin{align}\label{p}
\Psi(t)&=\frac{(s_2 +\mu+3\lambda)e^{s_2t}- (s_1
+\mu+3\lambda)e^{s_1t}}{s_2-s_1},\\
 \label{1a} s_{1,2}& = \frac{-\mu
-3\lambda \pm \sqrt{\mu^2 +6\mu\lambda +\lambda^2}}{2}.
\end{align}
And if $F(t)=1-e^{-\nu t}$, $\nu>0,\ t\ge 0,$ then
\begin{align}\label{pp} \V{}{\tau>t } &
 =\frac{\lambda(\nu-\mu)}{(\lambda +\nu)(\nu -\mu -2\lambda)+2\lambda^{2}}e^{-(\nu+\lambda)t}\notag\\
  &  +\frac{\nu}{s_2-s_1}\Bigl(\frac{s_2 +\mu+3\lambda}{\lambda +\nu+s_2}e^{s_2t}-
\frac{s_1 +\mu+3\lambda}{\lambda +\nu+s_1}e^{s_1t}\Bigr).
\end{align}
{\bf Proof.}\\
Consider first the simple case when  only one server is present and
let $P(t)$ denote the probability of its being in
 working condition at time $t.$ The memoryless property of
 an exponential distribution allows us to write down the following
 equation
\begin{align*}P(t) = e^{-\lambda t}+\lambda \mu \i{0}{t}e^{-\lambda u}\i{0}{t-u}e^{-\mu v}P(t-u-v)dvdu, \quad t>0
\end{align*}
which implies
\begin{align}\label{2}
P(t) =\frac{1}{\mu+\lambda}\Bigl( \mu +\lambda e^{-(\lambda+\mu)
t}\Bigr) , \quad t\ge 0.
\end{align}
Now we consider the original system but without delay, i.e.,
$\zeta\equiv 0,$ and let $\Psi(t)$ denote the probability that
such a system will be work without  breakage up to time $t$. Using
the memoryless property of exponential distributions of the random
variables $\xi_i$, $\eta_i,$ $i=1,2,$  and the formula (\ref{2}),
we get
\begin{align}\label{3}
\Psi(t)& = e^{-\lambda t}+\frac{\lambda\mu}{\lambda+\mu}\times\\
&
 \i{0}{t}e^{-\lambda u}(\mu +\lambda e^{-u(\lambda+\mu)})\i{0}{t-u}e^{-(\mu+\lambda)
v}\Psi(t-u-v)dvdu\notag\\
&+\frac{\lambda}{\lambda+\mu}
 \i{0}{t}e^{-\lambda u}(\mu +\lambda e^{-u(\lambda+\mu)})e^{-(\mu+\lambda)(t-u)}
du,
 \quad t>0.\notag
\end{align}
To find the function $\Psi(t)$ we use the method of Laplace
transform. If we  denote
\begin{align*}
\psi(s) =  \i{0}{\infty}e^{-st}\Psi(t)dt,\quad s>0,
\end{align*}
and apply to  (\ref{3}) the Laplace transform with a parameter
$s>0,$ we get
\begin{align*}
{}&\psi(s) =\frac{1}{s+\lambda} +\frac{\lambda\mu}
 {\mu+\lambda}\bigl( \frac{\mu}{s+\lambda}
 + \frac{\lambda}{s+\mu+2\lambda}\bigr)\frac{1}{s+\mu+\lambda}\psi(s)
 +\\
{}&+ \frac{\lambda}{\mu+\lambda}\frac{1}{s+\mu+\lambda} \bigl(
\frac{\mu}{s+\lambda}
+\frac{\lambda}{s+\mu+2\lambda}\bigr),\end{align*}
 which gives
\begin{align}\label{4}
\psi(s)& =\frac{s+\mu +3\lambda}{s^2+s(\mu+3\lambda)+2\lambda^2}=\notag\\
& =\frac{1}{s_2-s_1}\Bigl(\frac{s_2
+\mu+3\lambda}{s-s_2}-\frac{s_1 +\mu+3\lambda}{s-s_1}\Bigr),
\end{align}
where the numbers $s_{1,2}$ are defined in (\ref{1a}). Taking the
inverse Laplace transform, we obtain (\ref{p}). If now a delay is
present and its distribution is $\V{}{\zeta <t} =F(t)$, then we
obtain (\ref{5}). {\bf End of Proof.}

\section{Acknowledgments}
The authors are grateful to Prof. Steven Sine for the generous gift
of ACh receptor subunit cDNAs, Prof. Adam Szewczyk for the mito-GFP
construct. E.N. thanks Prof. Andrzej Majhofer for a suggestion to perform the Monte Carlo simulations for the dwell-time distributions.

\newpage
\onecolumngrid
\begin{center}
\begin{figure*}\includegraphics[width=15cm]{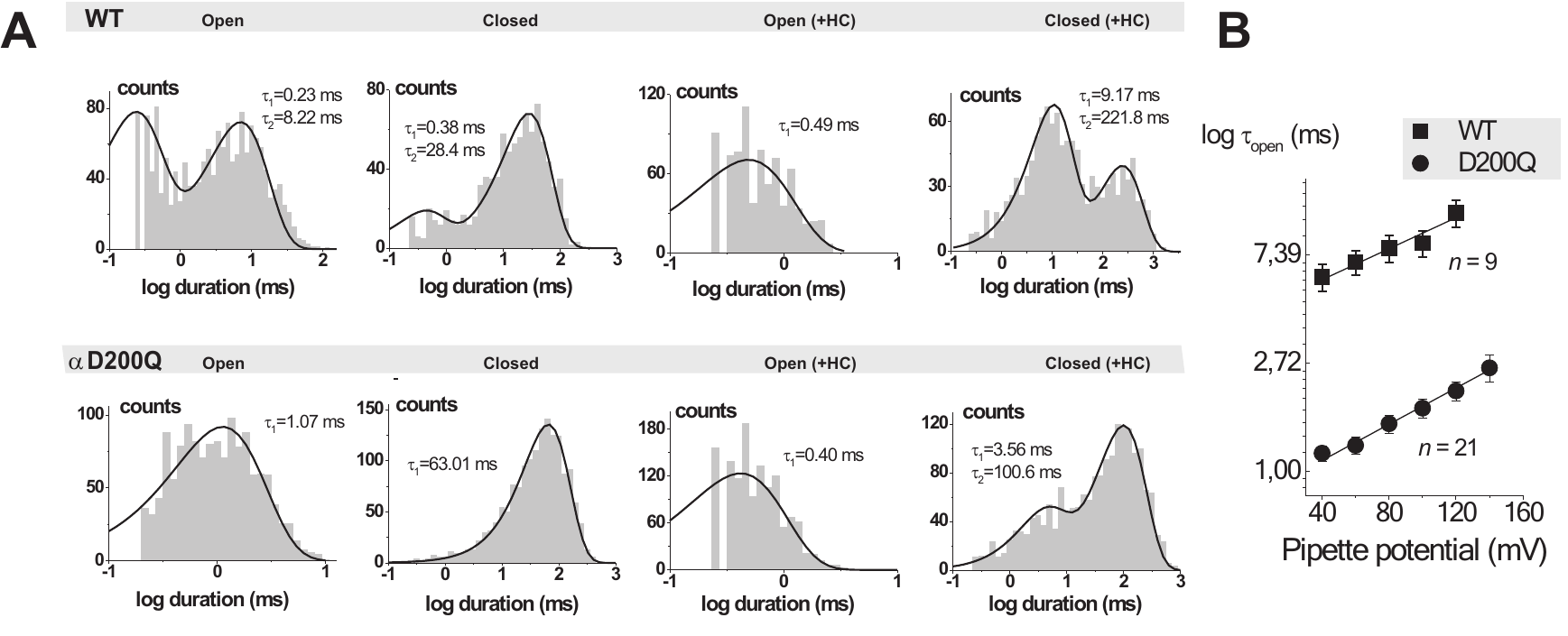}
\caption{ (A)Dwell-time histograms obtained from single-channel
recordings from WT and $\alpha$D200Q receptors at +60 mV of pipette
potential in control conditions and in the presence of 0.5 mM HC fitted
with a single or a double exponential function with time constants $\tau_{i}$;
ACh concentration was 500 nM in recordings from WT and 100 $\mu$M
from $\alpha$D200Q receptors. (B) Voltage dependency of the $\tau_{open}$ of the WT and
$\alpha$D200Q receptors. Data points represent the mean $\pm$ SEM
from $\it{n}$ membrane patches. Points are plotted on a
logarithmic scale and fitted with a function $f = A\exp((V-60)/b)$
where $V$ is a pipette potential,  $A$ = 6.93 ms, $b$ = 139 mV for
the WT receptor and $A$ = 0.8 ms, $b$ = 119 mV for  $\alpha$D200Q
receptor; note the similar slope (similar voltage-dependency) of
both plots. }\label{11}
\end{figure*}
\end{center}
\begin{figure}\includegraphics[width=8.0cm]{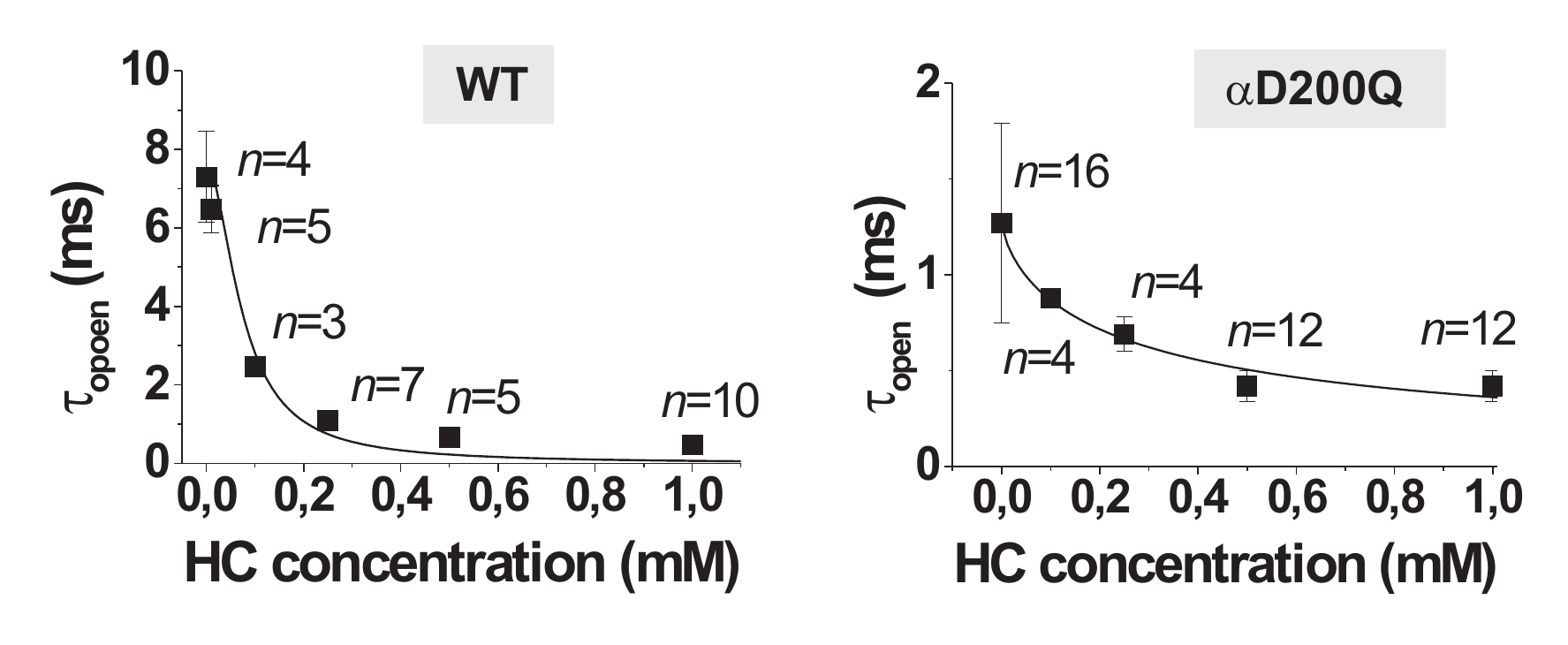}
\caption{HC decreased the open-channel lifetime. Plots show the mean
$\tau_{open}$ $\pm$ SEM from $\it{n}$ patches at +60 mV of pipette
potential; ACh concentration was 500 nM in recordings from WT and
100 $\mu$M from $\alpha$D200Q receptors. Solid line points are
fitted with a function $f =  1/(\alpha+ k_{+b}[x]^h)$ where $x$ is
HC concentration; $\alpha$ = 0.13 ms$^-1$,  $k_{+b} = $15.54
mM$^{-h}$ ms$^{-1}$, $h$ = 1.84; in WT receptor; $\alpha$ = 0.78
ms$^{-1}$, $k_{+b} = $1.98 mM$^{-h}$ms$^{-1}$, $h$ = 0.7 in
$\alpha$D200Q receptor. }\label{33}
\end{figure}

\begin{figure}\includegraphics[width=9.0cm]{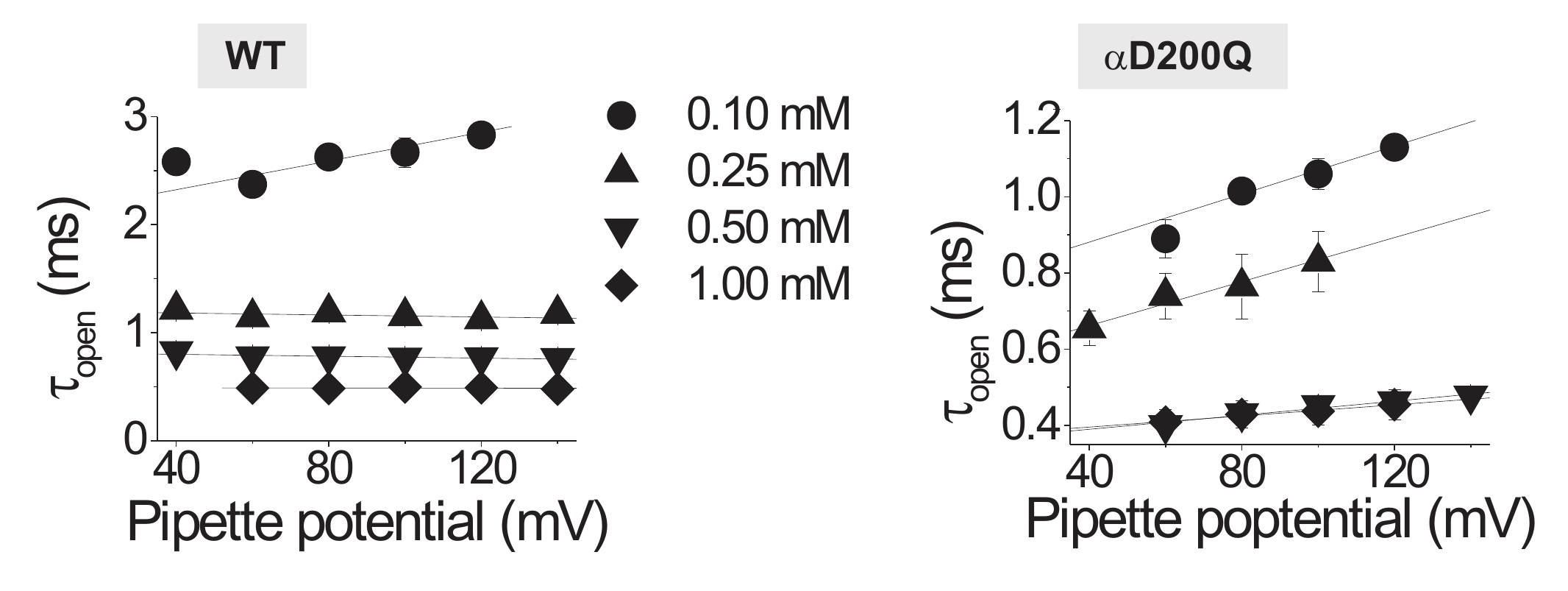}
\caption{The mean $\tau_{open}$ $\pm$ SEM at different membrane potentials and different HC concentrations.
 Solid lines are determined by a linear regression; $\it{n}$ - from 2 to 7;}\label{22}
\end{figure}

\begin{figure}\includegraphics[width=9.0cm]{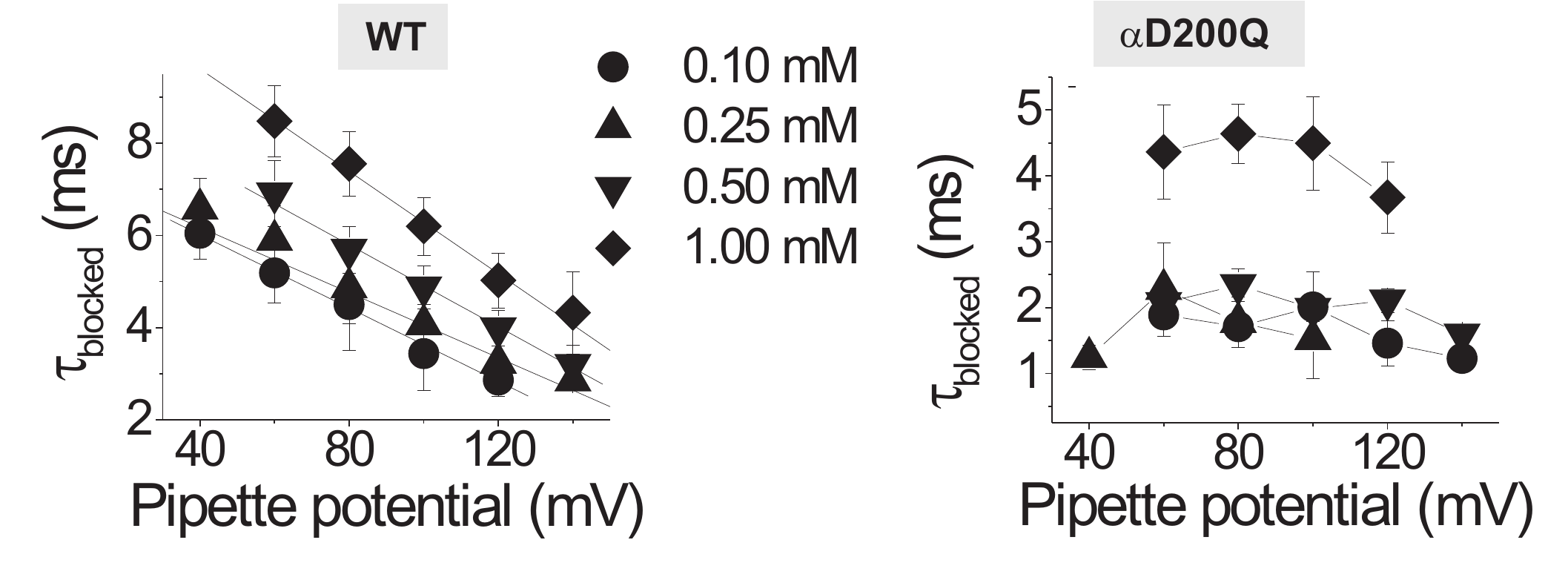}
\caption{ The mean $\tau_{blocked}$ $\pm$ SEM at different membrane
potentials and different HC concentrations; $\it{n}$ - from 2 to 7;
WT receptor: each solid line determined by a linear regression; the
slope significantly differs from 0 for all HC concentrations
(\textit{P}$ < $0.01); $\alpha$D200Q receptor: for [HC]$ < $ 1 mM
any clear relationship of the mean  $\tau_{blocked}$ with membrane
potential increase is not present. }\label{44}
\end{figure}

\begin{figure}\includegraphics[width=8.0cm]{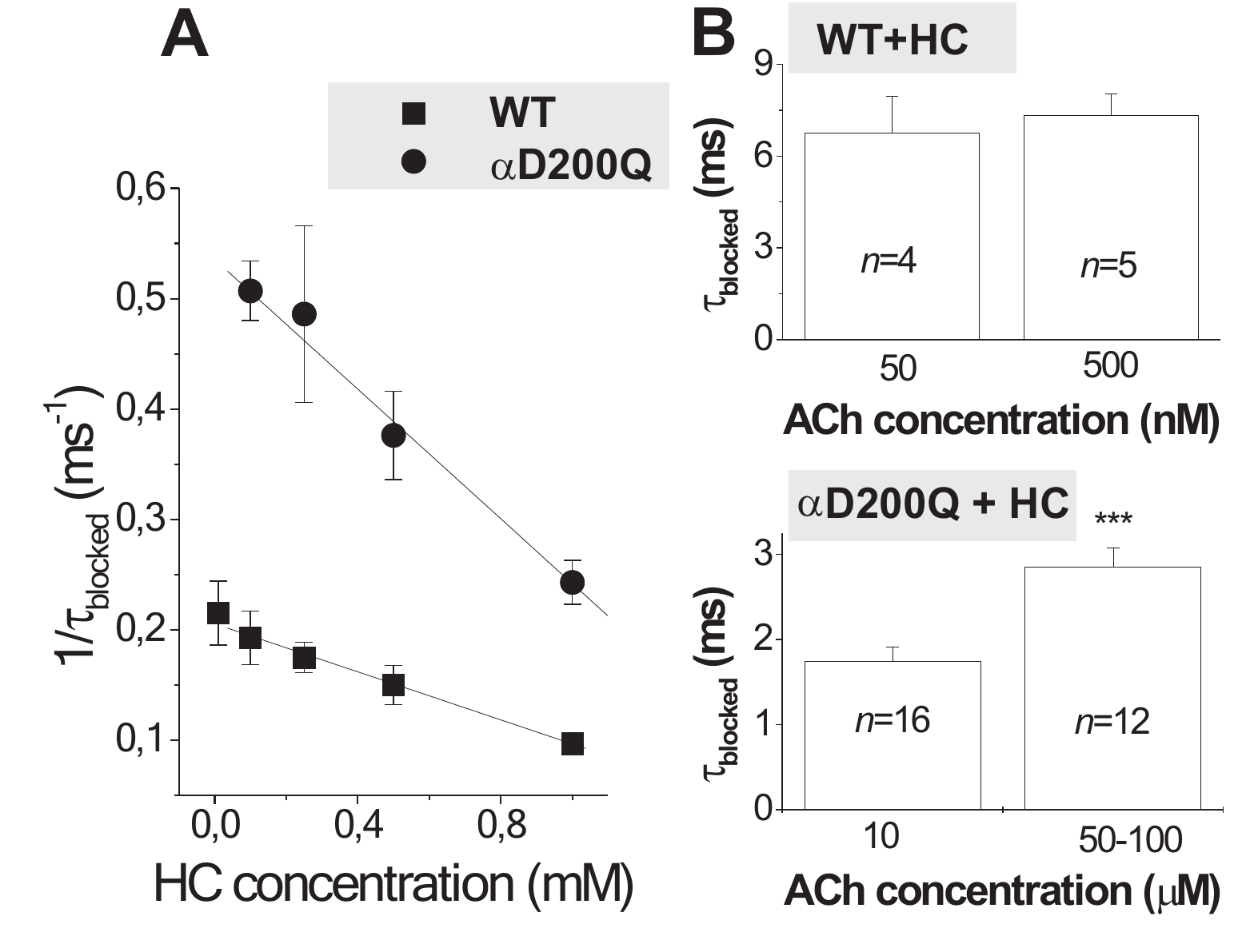}
\caption{(A) The reciprocal of the mean $\tau_{blocked}$ $\pm$ SEM
at +60 mV of pipette potential at different HC concentrations;
$\it{n}$ - the same as in Fig. 2; ACh concentration was 500 nM in
recordings from WT and 100 $\mu$M from $\alpha$D200Q receptors; each
line determined by a linear regression. (B) The mean
$\tau_{blocked}$ $\pm$ SEM in the presence of 0.5 mM HC ($\it{n}$ -
number of cells); $^{***}$ $\it{P}$ $ < $0.001.}\label{55}
\end{figure}

\begin{figure}\includegraphics[width=8.0cm]{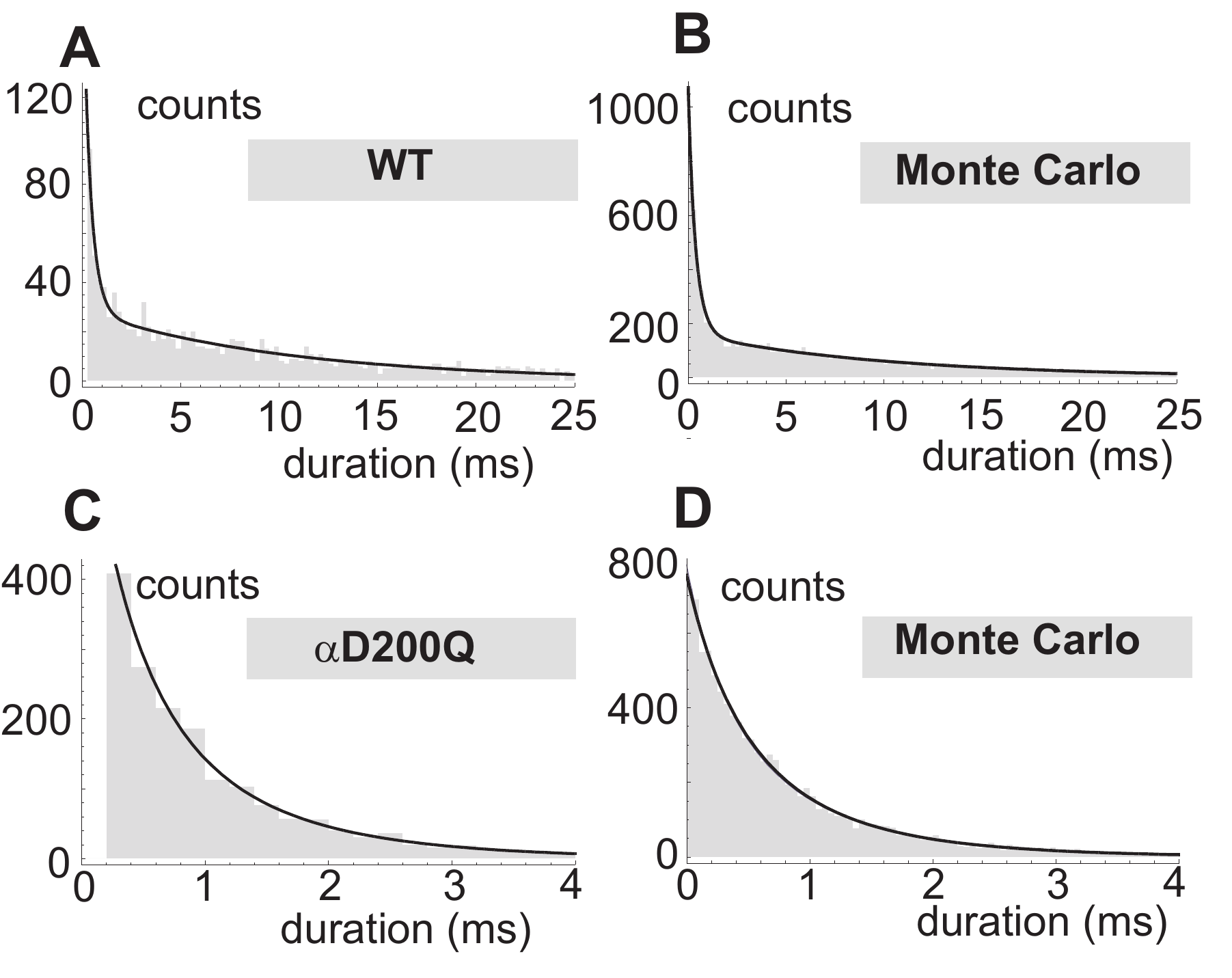}
\caption{(A) The example of the experimentally obtained open time
distribution of WT receptor fitted with the derivative of the
function (\ref{pp}). Fitting parameters are: $\lambda$ = 0.529
ms$^{-1}$, $\mu$  = 4.343 ms$^{-1}$,  $\nu$ = 1.986 ms$^{-1}$. The
function is: 355.8(0.39${\rm e}^{-2.5 t}$ + 0.19(0.30${\rm e}^{-5.84
t}$ + 0.41${\rm e}^{-0.09 t}$)). (B) The example of the distribution
obtained by Monte Carlo simulation of the 'oscillating' model
presented in the paper with the same parameters as in A fitted with
the derivative of the function (\ref{pp}); simulation performed for
10000 events; fitting parameters are: $\lambda$ = 0.540 ms$^{-1}$,
 $\mu$ = 4.409 ms$^{-1}$ and $\nu$ = 1.903 ms$^{-1}$. (C) The example of the
experimentally obtained open time distribution of $\alpha$D200Q
receptor fitted with the derivative of the function (\ref{pp}).
Fitting parameters are: $\lambda$ = 1.562 ms$^{-1}$,  $\mu$ = 1.624
ms$^{-1}$, $\nu$ = 0.648 ms$^{-1}$. The function is: 464.8(0.81${\rm
e}^{-2.2 t}$ + 0.70(0.31${\rm e}^{-5.4 t}$ + 0.77${\rm e}^{-0.90
t}$)). (D) The example of the distribution obtained by Monte Carlo
simulation of the 'oscillating' model presented in the paper with
the same parameters as in C fitted with the derivative of the
function (\ref{pp}); simulation performed for 10000 events; fitting
parameters are: $\lambda$ = 1.510 ms$^{-1}$,
 $\mu$ = 1.305 ms$^{-1}$ and $\nu$ = 0.473 ms$^{-1}$. }\label{66}
\end{figure}
\vspace {10pt}
\begin{figure}
\includegraphics[width=11.0cm]{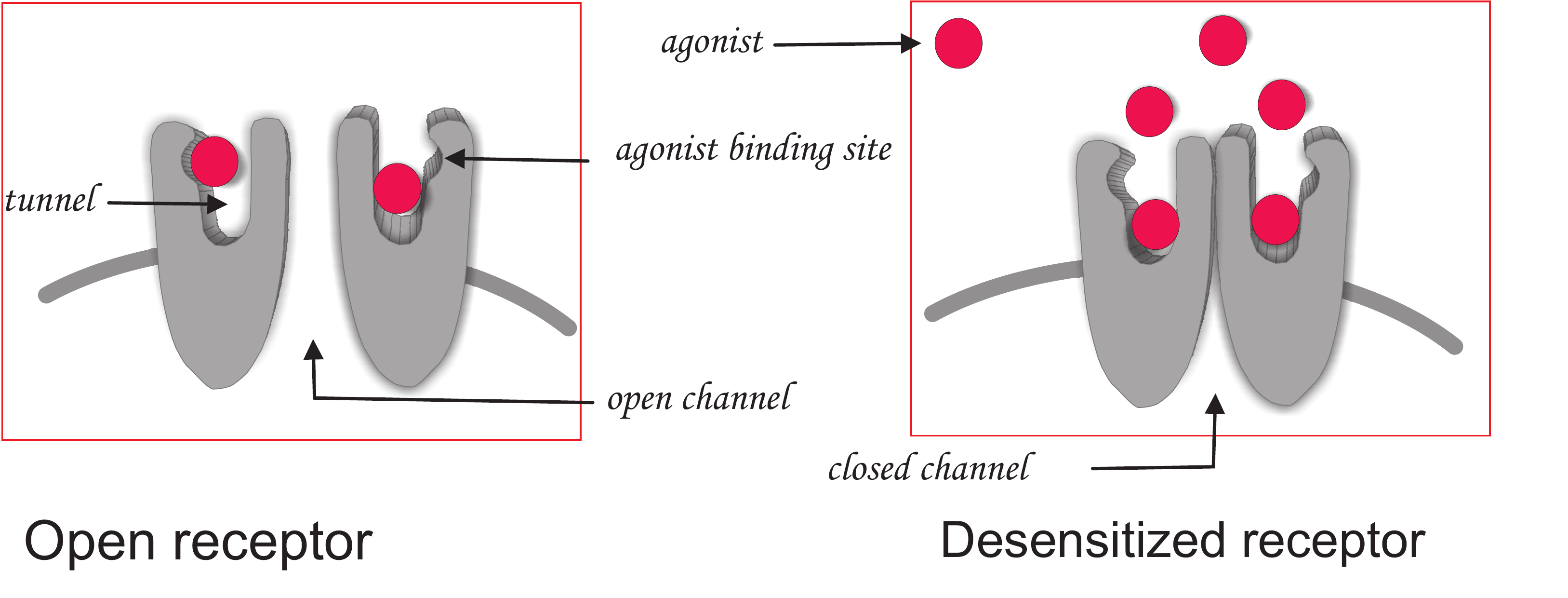}
\caption{Model of nACh receptor in two configurations: open and desensitized}
\end{figure}

\end{document}